\documentclass[aps,prd,twocolumn,superscriptaddress,amssymb,eqsecnum,showpacs,showkeyes,secnumarabic,graphics,floatfix,nofootinbib,tightenlines,longbibliography]{revtex4-1}

\usepackage{booktabs} % For better looking tables
\usepackage[margin=1in]{geometry} % To adjust margins
\usepackage{lipsum} % For dummy text

\usepackage{graphicx}
\usepackage{bm}
\usepackage{cancel}
\usepackage{epsfig}
\usepackage{dcolumn}
\usepackage{amsmath}
\usepackage{hhline}
\usepackage{enumerate}
\usepackage[utf8]{inputenc}
\usepackage[dvipsnames]{xcolor}
\usepackage[breaklinks=true,colorlinks=true,linkcolor=blue,urlcolor=Blue,citecolor=MidnightBlue,% PDF VIEW
bookmarks=true,bookmarksopenlevel=2]{hyperref}
\usepackage{bm}
\usepackage{amsmath}
\usepackage{amssymb}
\usepackage{subcaption}
\usepackage{academicons}

\begin{document}
\newcommand{\newc}{\newcommand}

\newc{\be}{\begin{equation}}
\newc{\ee}{\end{equation}}
\newc{\ba}{\begin{eqnarray}}
\newc{\ea}{\end{eqnarray}}
\newc{\bea}{\begin{eqnarray*}}
\newc{\eea}{\end{eqnarray*}}
\newc{\D}{\partial}
\newc{\ie}{{\it i.e.} }
\newc{\eg}{{\it e.g.} }
\newc{\etc}{{\it etc.} }
\newc{\etal}{{\it et al.}}
\newc{\lcdm}{$\Lambda$CDM }
\newc{\lscdm}{$\Lambda_{\rm s}$CDM }
\newc{\lcdmnospace}{$\Lambda$CDM}
\newc{\wcdm}{$w$CDM }
\newc{\plcdm}{Planck18/$\Lambda$CDM }
\newc{\plcdmnospace}{Planck18/$\Lambda$CDM}
\newc{\omom}{$\Omega_{0m}$ }
\newc{\omomnospace}{$\Omega_{0m}$}
\newcommand{\nn}{\nonumber}
\newc{\ra}{\Rightarrow}
\newc{\baodv}{$\frac{D_V}{r_s}$ }
\newc{\baodvnospace}{$\frac{D_V}{r_s}$}
\newc{\baodm}{$\frac{D_M}{r_s}$ }
\newc{\baoda}{$\frac{D_A}{r_s}$ } 
\newc{\baodanospace}{$\frac{D_A}{r_s}$}
\newc{\baodh}{$\frac{D_H}{r_s}$ }
\newc{\baodhnospace}{$\frac{D_H}{r_s}$}
\newc{\orcid}[1]{\href{https://orcid.org/#1}{\textcolor[HTML]{A6CE39}{\aiOrcid}}}
\newc{\ga}[1]{\textcolor{green}{[{\bf George}: #1]}}
\newc{\lk}[1]{\textcolor{orange}{[{\bf Lavrentios}: #1]}}
\newc{\lp}[1]{\textcolor{red}{[{\bf LP}: #1]}}
\newc{\snc}[1]{\textcolor{blue}{[{\bf Savvas}: #1]}}
\newc{\db}[1]{\textcolor{green}{[{\bf DB}: #1]}}

\title{Hubble tension tomography: BAO vs SnIa distance tension}

\author{Dimitrios  Bousis}\email{dimitriosbousis2002@gmail.com}
\affiliation{Department of Physics, University of Ioannina, GR-45110, Ioannina, Greece}

\author{Leandros Perivolaropoulos}\email{leandros@uoi.gr}
\affiliation{Department of Physics, University of Ioannina, GR-45110, Ioannina, Greece}

\date{\today}

\begin{abstract}
We investigate the redshift dependence of the Hubble tension by comparing the luminosity distances obtained using an up-to-date BAO dataset (including the latest DESI data) calibrated with the CMB-inferred sound horizon, and the Pantheon+ SnIa distances calibrated with Cepheids. Using a redshift tomography method, we find: 1) The BAO-inferred distances are discrepant with the Pantheon+ SnIa distances across all redshift bins considered, with the discrepancy level varying with redshift. 2) The distance discrepancy is more pronounced at lower redshifts ($z \in [0.1,0.8]$) compared to higher redshifts ($z\in [0.8,2.3]$). The consistency of \lcdm best fit parameters obtained in high and low redshift bins of both BAO and SnIa samples is investigated and we confirm that the tension reduces at high redshifts. Also a mild tension between the redshift bins is identified at higher redshifts for both the BAO and Pantheon+ data with respect to the best fit value of $H_0$ in agreement with previous studies which find hints for an 'evolution' of $H_0$ in the context of \lcdm.  These results confirm that the low redshift BAO and SnIa distances can only become consistent through a re-evaluation of the distance calibration methods. An $H(z)$ expansion rate deformation alone is insufficient to resolve the tension. Our findings also hint at a possible deviation of the expansion rate from the Planck18/$\Lambda$CDM model at high redshifts $z\gtrsim 2$. We show that such a deformation is well described by a high redshift transition of $H(z)$ like the one expressed by \lscdm even though this alone  cannot fully resolve the Hubble tension due to its tension with intermediate/low $z$ BAO data.
\end{abstract}
\maketitle

\section{Introduction}
\label{sec:Introduction}

The sound horizon scale $r_s$ at recombination (standard ruler)  and the standardized bolometric absolute magnitude $M_B$ of Type Ia supernovae (SnIa standard candles) have been used as probes for the measurement of cosmological distances and thus for the measurement of the Hubble constant $H_0$, the most fundamental parameter of cosmology. The best fit values of $H_0$ obtained using the two distance calibrators are at $5\sigma$ discrepancy with each other. This is the most important challenge\cite{Perivolaropoulos:2021jda,Abdalla:2022yfr,DiValentino:2021izs} for the current standard cosmological model the $\Lambda CDM$. 

%The assumptions behind each measurement and the classes of models.

Each type of measurement makes specific assumptions whose possible violation would lead to significant systematic errors that would invalidate the accuracy of the corresponding measurement. Measurements based on SnIa standard candles assume the validity of the distance ladder approach and in particular that physical laws and environmental effects around calibrated SnIa are the same in all the three rungs of the distance ladder. Measurements based on the sound horizon scale used as a standard ruler assume that the physical laws before recombination are consistent with the standard cosmological model and that the expansion history of the universe $H(z)$ is consistent with the standard \lcdm model and its parameters determined by the Planck18\cite{Planck:2018vyg} (\plcdm). 

In accordance with the above three types of assumptions there  are three classes of models for the resolution of the Hubble tension:
\begin{itemize}
\item
{\bf Ultralate time models:} These models\cite{Marra:2021fvf,Alestas:2020zol,Perivolaropoulos:2021bds}  assume that there is a physics or environmental change between the second and third rungs of the distance ladder ie that the Cepheid calibrated SnIa are not the same as the Hubble flow SnIa either due to a change of the physical laws or due to some change in their structure or environment (dust, metalicity etc). Thus, $M_B$ in the third rung of the distance ladder (Hubble flow rung $z\in [0.01,0.1]$) is lower than the value measured in the second rang. Thus these models explore the degeneracy between $H_0$ and $M_B$ in the context of the Hubble flow observable 
\begin{equation}
    {\cal M}\equiv M_B+5log\frac{c/H_0}{Mpc}+25
\label{calMdef}
\end{equation}
to lead to a lower $H_0$ measurement in the context of the distance ladder measurements.

The main problem of this class of models is fine tuning. There is no current clear theoretical motivation for the assumed physics transition at such ultralate redshifts ($z \sim 0.01$) even though it is straightforward to construct such models using a degree of fine tuning\cite{Perivolaropoulos:2022txg}. On the other hand, the physics transitions hypothesis is easily testable using a wide redshift data and in fact there are some hints for such an affect in both Cepheid and Tully-Fisser data \cite{Perivolaropoulos:2022khd,Alestas:2021nmi}.
\item
{\bf Early time models:} These models assume physics beyond the standard model to decrease the sound horizon scale at recombination
\begin{equation}
    r_s=\int_{z_{rec}}^{\infty}\frac{dz\; c_s(z)}{H(z;\Omega_{0b}h^2,\Omega_{0\gamma}h^2,\Omega_{0CDM}h^2)}
\end{equation}
to a lower value induced by Early Dark Energy \cite{Poulin:2018cxd,Kamionkowski:2022pkx,Simon:2022adh,Braglia:2020bym,Niedermann:2020dwg,Smith:2020rxx,Rezazadeh:2022lsf} (EDE), modified gravity\cite{Braglia:2020auw,Brax:2013fda,Abadi:2020hbr,Clifton:2011jh,Lin:2018nxe,DiValentino:2015bja,CANTATA:2021ktz,Rossi:2019lgt,Braglia:2020iik,FrancoAbellan:2023gec} or  dark radiation\cite{Seto:2021xua,Sakstein:2019fmf,Vagnozzi:2019ezj}. Thus, these models exploit the degeneracy of $r_s$ with $H_0$ in the context of the observable angular scale of the CMB sound horizon $r_s$
\begin{equation}
    \theta_s=\frac{r_s H_0}{\int_0^{z_{rec}}1/E(z)}
    \label{thetasdef}
\end{equation}
to lead to a higher value of $H_0$ while respecting the best fit \plcdm form of  $E(z)\equiv H(z)/H_0$ between recombination and present time.

The main problem of this class is that despite fine tuning of theoretical models that support them, they are only able to decrease the statistical significance of the Hubble tension and they can not fully eliminate it. In addition, they favor a higher value of matter density parameter $\Omega_{0m}$ thus worsening the $S_8$ tension \cite{Alestas:2021xes,Jedamzik:2020zmd,Vagnozzi:2023nrq,Vagnozzi:2021gjh}.
\item
    {\bf Late time models  ($H(z)$  deformation):} These models assume that there is deformation of the Hubble expansion history $H(z)$ with respect to the \plcdm prediction at late cosmological times ($z\lesssim 2$) \cite{Li:2019yem,Pan:2019hac,Panpanich:2019fxq,DiValentino:2019ffd,DiValentino:2019jae,Li:2020ybr,Clark:2020miy,DiValentino:2020kha}.  This deformation shifts the value of $H_0$ measured by the sound horizon scale standard ruler  to a value consistent with the distance ladder measurement by effectively changing the denominator of eq. (\ref{thetasdef}). However, as discussed in the present analysis even this model has difficulty to fit simultaneously BAO and SnIa data.

The main problem of this class of models is that $E(z)\equiv H(z)/H_0$ deformations are highly constrained by BAO and SnIa data in redshifts larger than $z\simeq 0.1$ \cite{Alestas:2021luu,Alestas:2021xes,Alestas:2020mvb,Brieden:2022lsd,Keeley:2022ojz,Clark:2020miy,DES:2020mpv,Anchordoqui:2022gmw,DES:2022doi,Cai:2022dkh,Heisenberg:2022gqk,Vagnozzi:2021tjv,Davari:2022uwd,DiValentino:2019jae,Gomez-Valent:2023uof}. Thus, this class of models have been put in disfavor during the last few years. Exception constitutes the \lscdm model which shifts the deformation to high redhifts ($z\simeq 2$) and makes it abrupt so that it has a minimal effect on late data while maintaining consistency with the distance to the CMB sound horizon at last scattering \cite{Akarsu:2022typ}.
\end{itemize}
%The H(z) deformation class of models

%Studies demonstrating the problems of the H(z) deformation class of models. The distance discrepancy issue.

The $H(z)$ deformation models are designed to be consistent with three types of cosmological data.

\begin{itemize}
    \item Low $z$ ($z\lesssim 0.1$) calibrated Hubble flow data (eg the SH0ES data \cite{Riess:2021jrx}) which measure $H_0= 73.04 \pm 1 km/sec Mpc^{-1}$.
    \item The distance to the CMB sound horizon scale $r_s$ as measured by the Planck CMB power spectrum.
    \item 
    The distance to BAO and SnIa data as obtained with the corresponding calibrators (CMB $r_s$ for BAO and Cepheid $M_B$ for SnIa).
\end{itemize}

These models perform very well in fulfilling the first two requirements. However, they all have problems fulfilling the third requirement. An interesting questing thus emerges 

{\it Is this difficulty due to the fact that there is a specific feature missed by the $H(z)$ parametrizations considered so far? Or there is a generic inconsistency between the BAO and SnIa data when the proper calibration is used in each case?}.

This  question is addressed in the present analysis using a redshift tomography method. We thus test the consistency of the cosmological distances obtained with the properly calibrated BAO and SnIa data with each other and with the \plcdm $E(z)$ in various redshift bins. We thus identify those redshifts where the distances obtained with the two classes of data are consistent and search for redshifts where there is a generic inconsistency between these distances. For the later redshift range it would be impossible to fit simultaneously the BAO and SnIa for any parametrization $H(z)$ due to the generic inconsistency between the two classes of data.

%Previous studies, promising approaches (LsCDM) and problems  (BAO data inconsistent with Cepheid $M_B$ and SnIa).

We thus find the level of inconsistency between $r_s$ calibrated BAO distances and Cepheid $M_B$ calibrated SnIa measured distances as a function of redshift. This may be interpreted as {\it the level of the Hubble tension as a function of redshift}. We also find the level of consistency of the measured distances with the predictions of \plcdm as a function of redshift.

Recent studies have investigated the redshift dependence of the Hubble tension using various probes and methods \cite{Dainotti:2021pqg,Colgain:2022nlb,Krishnan:2022oxe,Jia:2022ycc,Krishnan:2020obg,Krishnan:2020vaf}. These analyses consistently find a reduction in the tension at higher redshifts, hinting at a possible low-redshift origin for the discrepancy. However, the specific redshift range and significance of this effect varies between studies, motivating further investigation.

Previous studies \cite{Pogosian:2021mcs} have demonstrated a general trend for inconsistency between the cosmological distances obtained with BAO and SnIa data. We quantify this finding using redshift tomography and a more extensive up-to-date BAO dataset. Promising approaches like \lscdm \cite{Akarsu:2019hmw,Akarsu:2022typ} are also investigated, which introduce a rapid transition in $H(z)$ at high redshifts to improve consistency with SnIa data. However, we find that the fundamental tension between BAO and SnIa data creates problems even for these models.

%Structure of paper: ΙΙ Method: Binning distance measurements with Pantheon+ and BAO  (show BAO data Table). III Results: The Hubble tension in Redshift Bins , IV Conclusion, Summary, Future prospects.

Potential systematic effects that could influence the redshift dependence of the Hubble tension are an important consideration. For example, selection effects in the SnIa samples at different redshifts \cite{Brout:2018jch} or evolution in the SnIa luminosity with redshift \cite{Tutusaus:2018ulu} could impact the tension measurement. On the BAO side, the determination of the sound horizon scale $r_d$ relies on assumptions about early universe physics that could potentially be violated \cite{Jedamzik:2020krr}. While a full investigation of these effects is beyond the scope of this paper, they are important caveats to keep in mind when interpreting the results.

The structure of this paper is the following: In section II we present the method used in our analysis which involves binning distance measurements using BAO data and comparing them with the corresponding distances obtained with binned Pantheon+ SnIa data. This tomography is used for estimating the statistical level of the calibrator mismatch as a function of redshift (Hubble tension tomography). Our results are presented in section III. Finally, section IV we conclude, summarize and discuss possible extensions of our analysis.

\section{Method: Binning distance measurements with Pantheon+ and BAO}
\label{sec:Method}

The Pantheon+ sample provides SnIa luminosity distance $d_L$ and distance moduli $\mu$ measurements for redshifts in the range $z\in [0.001,2.3]$. These measurements are calibrated by the second rung of the distance ladder using Cepheids measuring the SnIa absolute magnitude as $M_B=-19.25\pm 0.01$ \cite{Riess:2021jrx,Perivolaropoulos:2022khd,Perivolaropoulos:2023iqj}. 

Assuming validity of the distance duality relation $d_L=(1+z)^2 D_A$, the BAO data calibrated by the comoving CMB sound horizon scale standard ruler $r_d=147.1Mpc$ at the end of the baryonic drag epoch \cite{Planck:2018vyg},  also lead to measurements of the luminosity distance $d_L$ and distance moduli $\mu$, in the redshift range $z\in [0.1,2.4]$. The CMB-inferred sound horizon scale $r_d$ is determined by fitting the standard \lcdm model to the Planck CMB power spectra data, assuming standard pre-recombination physics \cite{Planck:2018vyg}. It corresponds to the comoving size of the sound horizon at the end of the baryonic drag epoch, when photon pressure no longer supports acoustic oscillations in the primordial plasma (see \cite{Aubourg:2014yra} for a review). The consistency level of these measurements with the corresponding measurements of the Pantheon+ sample as a function of redshift bins reflects the level of the Hubble tension in redshift space. The method to perform this Hubble tension tomography is described in this section.

\subsection{Binning the distance moduli residual of Pantheon+}
\label{sec:BinningPan}
The Pantheon+ sample\cite{Brout:2021mpj,Scolnic:2021amr} consists of a table of 1701 rows (plus a header) and 47 columns. A $1701 \times 1701$ covariance matrix is part of it and it represents the covariance between all the SnIa due to systematic and statistical uncertainties of the distance moduli. In this analysis the information we need to utilize from the sample is: 
\begin{itemize}
    \item{The Hubble diagram redshift (with CMB and peculiar velocity corrections) }
    \item{The corrected apparent magnitude of the SnIa as well as its uncertainty which includes errors due to peculiar velocity }
    \item{The distance moduli $\mu$ calculated by subtracting the absolute magnitude of the SnIa $M_{SH0ES}=-19.253$ from the apparent magnitude and also their uncertainties obtained through the diagonal of the covariance matrix }
    \item{The distance moduli  $\mu_{Ceph}$, as found by the SH0ES distance ladder analysis, of the SnIa in Cepheid hosts with their uncertainties being incorporated in the covariance matrix }
\end{itemize}
The data can be found in the github repository \href{https://github.com/PantheonPlusSH0ES/DataRelease/tree/main/Pantheon%2B_Data/4_DISTANCES_AND_COVAR}{in this url}.

The SnIa distance moduli of the Pantheon+ sample measured in the Hubble flow are used to constrain the luminosity distance $d_L$ through the relation
\be \label{mu}
\mu(z)=m(z)-M_B=5log(d_L(z)/Mpc)+25
\ee 
where the luminosity distance is connected with the Hubble expansion rate:
\be 
\label{dl}
d_L(z)=(1+z)c\int^z_0\frac{dz'}{H_0\;E(z')}
\ee 
$c$ is the speed of light and the dimensionless Hubble parameter $E$ in the context of \lcdm is given by:
\be \label{e}
E=\sqrt{(a^{-3}\Omega_{0m}\left(1+\frac{a_{eq}}{a}\right)+\left( 1-\Omega_{0m}(1+a_{eq}) \right)}
\ee 
with $a_{eq}$  the scale factor corresponding to the time when matter and radiation were equal 
\be \label{aeq}
a_{eq}=\frac{1}{(z_{eq}+1)}  ,  z_{eq}=2.5\cdot 10^4 \Omega_{0m}h^2(T_{CMB}/2.7)^{-4}
\ee 
with $T_{CMB}=2.7255K$.
%We are particularly interested in the distance moduli of the SnIa. The distance modulus for a SnIa at a given redshift is the difference between absolute luminosity $M$ and apparent luminosity $m_B$. Thus it can be written as:

To get the theoretical distance modulus corresponding to the \plcdm model we simply  need to incorporate the Planck 2018 values for the present matter density parameter $\Omega_{0m}=0.3166$ and the Hubble constant $H_0=67.4 \quad km \quad s^{-1} \quad Mpc^{-1}$ ($h=0.674$) into equations (\ref{mu})-(\ref{e}) to obtain $\mu_{P\Lambda CDM}(z)$. 

We define the distance modulus residual between the theoretical values of the \plcdm model and the observational data obtained through the Pantheon+ sample:
\be 
\Delta \mu(z_i)\equiv \mu_{obs}(z_i)-\mu_{P\Lambda CDM}(z_i)
\label{delmu}
\ee 
We have binned this residual in 13 evenly spaced bins for $z\in [0.001,2.3]$ to compare with corresponding BAO data. The redshift for each bin is the mean of all redshifts within said bin. For the calculation of the distance modulus for each bin the covariance matrix was incorporated. This was done by first constructing a vector for each bin $j$ with components 
\be \label{qi}
Q_i=\Delta \mu(z_i)-\mu_{cj}
\ee 
Components of $Q$ out of the $j$ bin were set to 0. In (\ref{qi}), $\mu_{cj}$ is the residual distance modulus in the $j^{th}$ bin and $\Delta \mu(z_i)$ is defined by eq. (\ref{delmu}) for a given redshift $z_i$ in the redshift range of the corresponding bin. Then to obtain the best fit value of $\Delta \mu(z_{Meanj})$ (with $z_{Meanj}$ being the mean redshift in the redshift range of the $j^{th}$ bin), we found the value of the best fit $j^{th}$ bin distance modulus residual  $\mu_{cj}$ that minimizes the appropriate $\chi^2$ function:
\be \label{qichi}
\chi^2_{j}=Q^T \cdot C^{-1} \cdot Q
\ee 
where $C^{-1}$ is the inverse of the covariance matrix of the Pantheon+ sample. Using the definition (\ref{qi}) in (\ref{qichi}) the proper part of the inverse covariance matrix corresponding to the $j^{th}$ bin is automatically selected. In this fashion we end up with 13 different \plcdm residual distance moduli values $\mu_{cj}$, one for each bin. The uncertainties for each bin $j$ were calculated in the standard way using the inverse of the Fisher matrix
\be 
F=\frac{1}{2}\frac{\partial^2\chi^2(\mu_{cj})}{\partial\mu_{cj}^2}
\ee 
We have also split  the Pantheon+ sample into two larger bins for redshifts $0.1<z<0.8$ and $0.8<z<2.3$ in order to fit for each bin the \lcdm cosmological parameters and identify the self consistency of the Pantheon+ sample and also compare with the corresponding best fit parameters of a BAO sample discussed in the next subsection. For each bin a data vector can be constructed as follows:
\be 
Q'_i=\begin{cases} 
 m_{B,i}-M_B-\mu^{Ceph}_i &    , \text{Cepheid hosts} \\
      m_{B,i}-M_B-\mu_{\Lambda CDM}(z_i)       &   , \text{Otherwise} \\
      \end{cases}
\ee 
where in each bin we have also included the Cepheid host SnIa distance moduli as found by the SH0ES distance ladder analysis.  For each bin we evaluate the $\chi^2$ function:
\be 
\chi'^2_{bin}=Q'^T \cdot C^{-1} \cdot Q'
\ee 
and minimize it to obtain the best fit values for $M_B$, $\Omega_{0m}$ and $h$ and the corresponding likelihood contours discussed in the next section. We have also constructed the distance moduli residual for the two new large bins as described above for the smaller bins.  These distance moduli residuals and \lcdm parameter contours were then compared with the corresponding BAO quantities described in the next subsection.

\subsection{Distance moduli from BAO}
We have  compiled a list of transverse BAO measurements,  of the rescaled comoving angular diameter distance $\frac{D_M}{r_d}$, where $r_d$ is the sound horizon scale during the drag epoch in which the photon drag on regular matter (baryons) becomes negligible.  The sample created includes both data as recent as the DESI 2024 \cite{DESI:2024mwx}  as well as some earlier measurements including the latest releases of the Sloan Digital Sky Survey (SDSS) and the Dark Energy Survey (DES). The full sample we use is shown in Table \ref{Dmrd}

The data in Table \ref{Dmrd} are from anisotropic BAO analyses, they have been derived using a fiducial model to convert the observed data into physical distances, and then isolating the angular component ($ D_M $) to compute $ D_M / r_d $. The distinction between anisotropic BAO analyses and 2D BAO data may be described as follows:
\begin{itemize}
\item Anisotropic BAO analyses incorporate the full three-dimensional distribution of galaxies. These analyses take into account both the radial and transverse components simultaneously, often using a fiducial cosmological model to convert observed angles and redshifts into physical distances. This allows for a more detailed understanding of the BAO signal and its anisotropies, but it introduces model dependencies in the process.
\item  In 2D BAO analyses, the BAO signal is examined in a two-dimensional plane, often focusing separately on the transverse (angular) and radial (line-of-sight) directions. This method typically avoids using a fiducial cosmological model to convert angular separations and redshifts directly into physical distances. Instead, it keeps the analysis in observational units (angles and redshifts), which can then be used to infer cosmological parameters in a somewhat model-independent way. This approach is designed to minimize assumptions about the underlying cosmology when extracting the BAO signal.
\end{itemize}

\begin{table*}
\centering
\caption{$D_M/r_d$ measurements and the corresponding references. The sample consists of 33 measurements. Measurements with an asterisk are the $D_M/r_d$ value calculated using the angular diameter distance $D_A=D_M/(1+z)$ and the fiducial $r_d^{fid}$ value $r_d=147.18 Mpc$ obtained by Planck18 in the context of \lcdm. More details about the construction of the * datapoints are presented in the Appendix. \ref{Appendix}.}
\label{Dmrd}
\begin{tabular}{|lllll|}
\hline
N& \textbf{$z_{eff}$} & \textbf{$D_m/r_d$} & \textbf{References}                                                                 & \textbf{Year} \\ \hline

1&0.32*              & 8.54±0.25          & Chuang et al   \cite{BOSS:2016goe}                                         & 2016          \\
2& 0.32               & 8.76±0.14          & BOSS collaboration  \cite{BOSS:2016wmc}                                             & 2016          \\

3&0.38               & 10.27±0.15         & BOSS collaboration   \cite{BOSS:2016wmc}                                               & 2016          \\
%0.39              & 10.27±0.15         & BOSS collaboration\cite{BOSS:2016wmc}                                        & 2017          \\
4&0.44*               & 11.79±1.11         & Blake et al    \cite{Blake:2012pj}                              & 2012          \\
5&0.51               & 13.38±0.18         & BOSS collaboration \cite{BOSS:2016wmc}                                       & 2016          \\
%0.51               & 13.38±0.18         & BOSS collaboration        \cite{Alam_2017}                                          & 2016          \\
%Can't find it% 0.51               & 11.92±0.18         & Carvalho et al  \cite{Carvalho:2015ica}                                                & 2015          \\
6&0.51               & 13.62±0.25         & DESI Collaboration  \cite{DESI:2024mwx}                                                & 2024          \\
7&0.54*               & 14.76±0.68         & Seo et al     \cite{Seo:2012xy}                                                       & 2012          \\
8&0.57               & 14.74±0.24         & BOSS collaboration   \cite{BOSS:2016wmc}                                               & 2016          \\
9&0.59*               & 15.29±0.25          & Chuang et al   \cite{BOSS:2016goe}                                         & 2016          \\
10&0.60*                & 14.99±1.04         & Blake et al       \cite{Blake:2012pj}                           & 2012          \\
11&0.61               & 15.45±0.22         & BOSS collaboration    \cite{BOSS:2016wmc}                                              & 2016          \\
%0.69               & 17.48±0.23         & Nadathur et al  \cite{Nadathur_2020}                                                & 2020          \\
12&0.70*              & 17.28±0.89         & Sridhar et al  \cite{Sridhar:2020czy}                                                  & 2020          \\
%0.70               & 17.65±0.30         & Bautista et al\cite{eBOSS:2020lta}, Gil-Marin et al\cite{eBOSS:2020hur} & 2020          \\
13&0.70                & 17.96±0.51         & Zhao Gong-bo et al \cite{eBOSS:2020rpt}                                                 & 2021          \\
14&0.71               & 16.85±0.32         & DESI Collaboration  \cite{DESI:2024mwx}                                                & 2024          \\
15&0.73*               & 18,03±1.26         & Blake et al  \cite{Blake:2012pj}                                & 2012          \\
16&0.77               & 18.85±0.38         & Wang et al    \cite{eBOSS:2020xwt}                                                      & 2020          \\
17&0.80               & 19.54±2.07         & Zhu et al       \cite{eBOSS:2018apc}                                        & 2018          \\
%can't find it% 0.81               & 19.46±0.78         & Abbot et al       \cite{DES:2017txv}                                      & 2017          \\
%0.84              & 18.92±0.51         & DES collaboration \cite{DES:2021esc}                                                                   & 2021          \\
18&0.85              & 19.51±0.41         & DES collaboration     \cite{DES:2024cme}                                                     & 2024          \\
%0.85              & 18.90±0.78         & Zhao Gong-bo et al   \cite{Zhao_2021}                                               & 2021          \\
%0.85              & 19.84±0.53         & Chan et al    \cite{Chan_2022}                                                      & 2022          \\
19&0.85              & 19.5±1.0           & Tamone et al\cite{eBOSS:2020qek}, de Mattia et al\cite{eBOSS:2020fvk}   & 2020          \\
20&0.87*               & 21.39±1.39         & Sridhar et al   \cite{Sridhar:2020czy}                                                 & 2020          \\
21&0.93               & 21.71±0.28         & DESI Collaboration  \cite{DESI:2024mwx}                                                & 2024          \\
22&1.00               & 23.13±2.07         & Zhu et al    \cite{eBOSS:2018apc}                                           & 2018          \\
23&1.32               & 27.79±0.69         & DESI Collaboration  \cite{DESI:2024mwx}                                                & 2024          \\
24&1.48               & 30.21±0.79         & Hou et al\cite{eBOSS:2020gbb}, Neveux et al\cite{eBOSS:2020uxp}         & 2020          \\
25&1.50               & 30.51±1.85         & Zhu et al       \cite{eBOSS:2018apc}                                        & 2018          \\
26&2.00               & 36.18±1.69         & Zhu et al       \cite{eBOSS:2018apc}                                        & 2018          \\
27&2.20               & 38.09±1.73         & Zhu et al       \cite{eBOSS:2018apc}                                        & 2018          \\
28&2.33               & 37.5±1.1           & Du Mas des Bourbuox et al\cite{eBOSS:2020tmo}                                 & 2020          \\
29&2.33               & 39.71±0.94         & DESI Collaboration  \cite{DESI:2024mwx}                                                & 2024          \\

30&2.34*               & 37.72±2.18         & Delubac et al  \cite{BOSS:2014hwf}                                                  & 2014          \\
31&2.35               & 36.3±1.8           & Blomqvist et al    \cite{eBOSS:2019qwo}                                            & 2019          \\
%2.36               & 36.29±1.34         & Font-Ribera et al    \cite{Font_Ribera_2014}                                        & 2014          \\
%2.40               & 36.6±1.2           & Abbott et al    \cite{DES:2017txv}                                        & 2018          \\
32&2.40               & 36.6±1.2           & De Mas des Bourboux et al \cite{BOSS:2017uab}                           & 2017          \\
 \hline
\end{tabular}
\end{table*}

The distance modulus residual can now be written as:
\be \label{mures}
\begin{split}
\Delta \mu(z_i) & =\mu_{obs}(z_i)-\mu_{P\Lambda CDM}(z_i) \\
& =5log \left( \frac{d_L^{obs}(z_i)}{d_L^{P\Lambda CDM}(z_i)} \right)
\end{split}
\ee
The ratio of the observed luminosity distance over the Planck 2018 theoretical luminosity distance, $\frac{d_L^{obs}(z_i)}{d_L^{P\Lambda CDM}(z_i)}$, can be replaced by:
\be 
\frac{d_L^{obs}(z_i)}{d_L^{P\Lambda CDM}(z_i)}=\frac{D_M^{obs}(z_i)}{D_M^{P\Lambda CDM}(z_i)}
\ee 
since the comoving angular distance $D_M$ is simply $D_M=(1+z)D_A$, where $D_A$ is the angular diameter distance which in turn is $D_A=d_L/(z+1)^2$.
Therefore, the distance moduli residuals can be written as:
\be 
\Delta \mu (z_i)=5log \left( \frac{D_M^{obs}(z_i)}{D_M^{P\Lambda CDM}(z_i)} \right)
\ee 
where we have estimated $D_M^{obs}(z_i)$ by multiplying the entries of Table \ref{Dmrd} by the best fit value $r_d=147.18Mpc$ from Planck18\cite{Planck:2018vyg}. For the theoretical \plcdm $D_M(z_i)$ we have incorporated the Planck 2018 \lcdm values for the Hubble constant and the present matter density parameter as mentioned above in section \ref{sec:BinningPan}. We also assume flat universe and thus use the present $\Lambda$ density parameter $\Omega_{0\Lambda}=0.685$, the present radiation density parameter $\Omega_{0r}=9.26*10^{-5}$ and the sound horizon drag scale $r_d=147.18$:
\begin{multline}
D_M^{P\Lambda CDM}(z_i)= \\ \int ^{z_i}_0 \frac{c}{H_0 \sqrt{\Omega_{0m}(1+z)^3 +\Omega_{0\Lambda}+\Omega_{0r}(1+z)^4}}
\end{multline} 
The distance modulus residual $\Delta \mu (z_i)$ can then be plotted as a function of redshift $z_i$  (left panel of Fig. \ref{fig1}). The uncertainties are calculated using the error propagation method as follows:
\be \label{UncDm}
\sigma_i=\frac{5}{ln10}\frac{1}{D_M^{obs}(z_i)} \cdot \sigma_{D_M^{obs}}(z_i)
\ee
where we have omitted the effect of the uncertainty of $r_d$ because it is easy to show that it is subdominant. We have also merged some of the points which have the same effective redshift value. This was done by first calculating the weighted $D_M$ ratios \cite{press2007numerical}:
\be 
D_{M,weighted}=\frac{\sum D_{Mi}/\sigma_i}{\sum 1/\sigma_i}
\ee 
and then calculating the combined uncertainty:
\be 
\sigma_{combined}=\frac{1}{\sqrt{\sum 1/\sigma_i}}
\ee 
 For the construction of the right panel of Fig. \ref{fig1}, we have also considered three redshift bins for redshift ranges $0.1<z<0.8$, $0.8<z<2.0$ and $2.0<z<2.5$. For each bin we get the distance modulus residuals $\Delta\mu_i$ as shown in Eq. (\ref{mures}) and the uncertainties $\sigma_i$ for each residual (see Eq. \ref{UncDm}). The redshift for each bin is the mean of all the redshifts of said bin, the distance modulus residual is the weighted mean:
\be 
\Delta\mu_{bin}=\frac{\sum \frac{\Delta\mu_i}{\sigma^2_i}}{\sum 1/\sigma_i^2}
\ee 
and the uncertainty is the combined uncertainty:
\be 
\sigma_{bin}= \sqrt{\frac{1}{\sum 1/\sigma_i^2}}
\ee 
The bin residulal distance moduli are shown in the right panel of Fig. \ref{fig1} along with the two Pantheon+ bins discussed at the end of section \ref{sec:BinningPan} for a comparison to be made. The comparison is discussed in the next section.

For the comparison of the best fit cosmological parameters $h$ and \omom with Pantheon+,   we have constructed two redshift bins in the same redshift ranges as those used for the Pantheon+ data ($0.1<z<0.8$ and $0.8<z<2.5$ \footnote{It should be noted here that the redshift range of the second bin goes beyond redshift 2.3, up to redshift 2.5 since we have BAO data for these redshifts as well.}. 

Thus, for each bin we construct the $\chi^2$ function as follows:
\be  
\chi^2=\sum \frac{(D_M^{obs}(z_i)-D_M^{\Lambda CDM}(z_i))^2}{\sigma_{D_M^{obs}}^2(z_i)+\sigma_{r_d}^2}
\label{chi2baoplcdmbin}
\ee 
where we estimate $(D_M^{obs}(z_i)$ using the Planck18 best fit value for $r_d=147.18Mpc$. By minimizing this function we obtain the best fit values for $h$ and $\Omega_{0m}$. 
The parameter uncertainties were again calculated using the Fisher matrix:
\be 
F_{ij}=\frac{1}{2}\frac{\partial^2\chi^2(p)}{\partial p_i\partial p_j}
\ee 
with the two parameters $p_i$ and $p_j$ being $\Omega_{0m}$ and $h$. The uncertainties in this case are the square roots of the diagonal elements of the inverse Fisher matrix. The corresponding contour plots can then be constructed to represent these results.

%The $H(z)$ deformation models are designed so that they predict the same angular diameter distance to the sound horizon scale at recombination ($z\simeq 1100$) as \plcdm (thus they are consistent with the CMB power spectrum) while at the same time they are consistent with the local measurements of $H_0$ \cite{Alestas:2020mvb}. Such models include the wCDM model with $w=-1.2$ \cite{Alestas:2020mvb} and $\Lambda_s$CDM with ... cite... The residual distance moduli of these models with respect to the distance modulus corresponding to \plcdm are shown in Fig. ... Notice the succesfull interpolation of wCDM between the low z SnIa calibrated with Cepheids and the distance to the CMB sound horizon at high z. The interpolation for \lscdm is less successfull as it is not fully consistent with the Cepheid calibration of the low z SnIa distance moduli. 

\subsection{Model Residuals for the $wCDM$ and \lscdm Models}

The study of $H(z)$ deformation models is driven by the need for a cosmological framework that is in agreement with Cosmic Microwave Background (CMB) observations at high redshifts ($z\simeq 1100$) as well as with local measurements of the Hubble constant ($H_0$). These models are designed to maintain the angular diameter distance to the sound horizon scale at recombination as observed by the \plcdm model, thereby ensuring consistency with the CMB power spectrum. They also assume that $\Omega_{0m} h^2$ has the same value as the one measured by the CMB angular power spectrum while setting constraining $h$ by local measurements \cite{Alestas:2020mvb}. Interesting such models are the $wCDM$ \cite{Alestas:2020mvb} and \lscdm\cite{Akarsu:2019hmw,Akarsu:2022typ} models investigated for their $H(z)$ deformation approach to the $H_0$ tension.

\textbf{wCDM Model:} The $wCDM$ model modifies the standard equation of state for dark energy, $w$, from the \plcdm value of $-1$. The Hubble parameter for this model is given by
\begin{equation}
\label{wcdmH}
\left(\frac{H}{H_0}\right)^2 = \Omega_{r0}(1 + z)^4 + \Omega_{m0}(1 + z)^3 + \Omega_{DE0}(1 + z)^{3(1+w)}.
\end{equation}

\textbf{$\boldsymbol{\Lambda sCDM}$ Model:} Alternatively, the \lscdm model posits a step-like alteration in the dark energy density at a transition redshift $z_*$. The corresponding Hubble parameter is modeled as
\begin{equation}
\left(\frac{H}{H_0}\right)^2 = \Omega_{r0}(1 + z)^4 + \Omega_{m0}(1 + z)^3 + \Omega_{\Lambda s0}\text{sgn}(z - z_*),
\end{equation}
where $\text{sgn}$ represents the signum function, indicative of the dark energy density change.

In this context, the residual distance modulus, \(\Delta \mu(z)\), quantifies the difference between the distance modulus as predicted by a given cosmological model, \(\mu_{\text{model}}(z)\), and the distance modulus according to the Planck18 \(\Lambda\)CDM standard model, \(\mu_{\text{\plcdm}}(z)\). It is defined as:
\begin{equation}
\Delta \mu(z) = \mu_{\text{model}}(z) - \mu_{\text{\plcdm}}(z).
\end{equation}
where 'model' may correspond eg to $wCDM$ or to \(\Lambda sCDM\).
This residual is a critical diagnostic tool to assess the viability of models such as $wCDM$ and \lscdm in resolving the \(H_0\) tension, by comparing their predictions with the well-established \lcdm  model as fitted to Planck data and with the intermediate SnIa and BAO data.

\begin{figure*}[ht!]
\centering
\includegraphics[width=\textwidth]{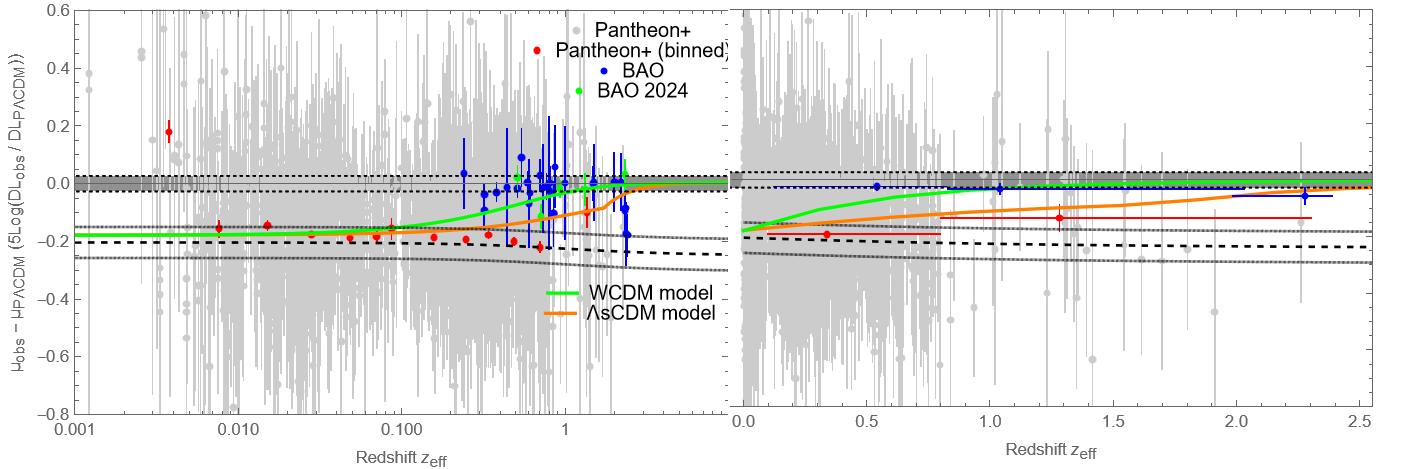}
\caption{The binned BAO and Pantheon+ luminosity distance moduli (residuals with respect to the \plcdm). The residual distance moduli for the best fit $wCDM$ and \lscdm are also shown. Larger redshift bins are shown in the right panel. The inconsistency between BAO and SnIa measured distance moduli is evident. The full Pantheon+ distance moduli are also shown in the grey background. The green points correspond to the latest DESI BAO data.}
\label{fig1}
\end{figure*}

For the $wCDM$ model, with $w\simeq -1.2$, there is a successful interpolation between the Cepheid-calibrated low-$z$ SnIa measurements and the CMB-inferred distances to the sound horizon at high-$z$ \cite{Alestas:2020mvb}. This is illustrated in Figure \ref{fig1}, where the residual distance moduli for the $wCDM$ model are compared to those of the \lscdm model, with the BAO and SnIa properly calibrated distance moduli residuals and with the SH0ES-Patheon+ \lcdm \cite{Brout:2022vxf} and \plcdm \cite{Planck:2018vyg} residual distance moduli (the later has central value at 0 by definition). The best fit \lscdm model\cite{Akarsu:2019hmw,Akarsu:2022typ,Akarsu:2023mfb} with  parameter values obtained from \cite{Akarsu:2023mfb} (Table I last column) ($z_*=1.72$), provides a better fit to the more constraining SnIa data at the expense of a worse fit to the BAO data which however have larger uncertainties and thus it leads to a better $\chi^2$ than \lcdm and other $H(z)$ deformation models like $wCDM$ as shown in Fig. \ref{fig1}. Thus this feature or high $z$ 'break' of $H(z)$ provides an advantage for \lscdm but does not overcome the fundamental problem of the $H(z)$ deformation models which is the generic inconsistency between SnIa and BAO data.

\textbf{Parameter values of $H(z)$ deformation models:} 
For the $wCDM$ model residual we use the cosmic parameters as calculated by the method described in \cite{Alestas:2020mvb}. For the $w$ parameter we use the following formula:
\be 
h(w) \simeq - 0.3093w + 0.3647
\ee 
where  for $h$ we set $0.73$ to find $w_{model} = - 1.188$. Then, for the present matter density parameter we use the \plcdm value
\be 
\overline{\omega_{0m}}=0.1430 \pm 0.0011
\ee 
where $\overline{\omega_{0m}}=\Omega_{0m} h^2$ which leads to $\Omega_{0m}=0.267$ for $H(z)$ deformation models consistent with local measurements of $h$ (eg $wCDM$).  

\begin{figure*}[ht!]
\centering
\includegraphics[width=\textwidth]{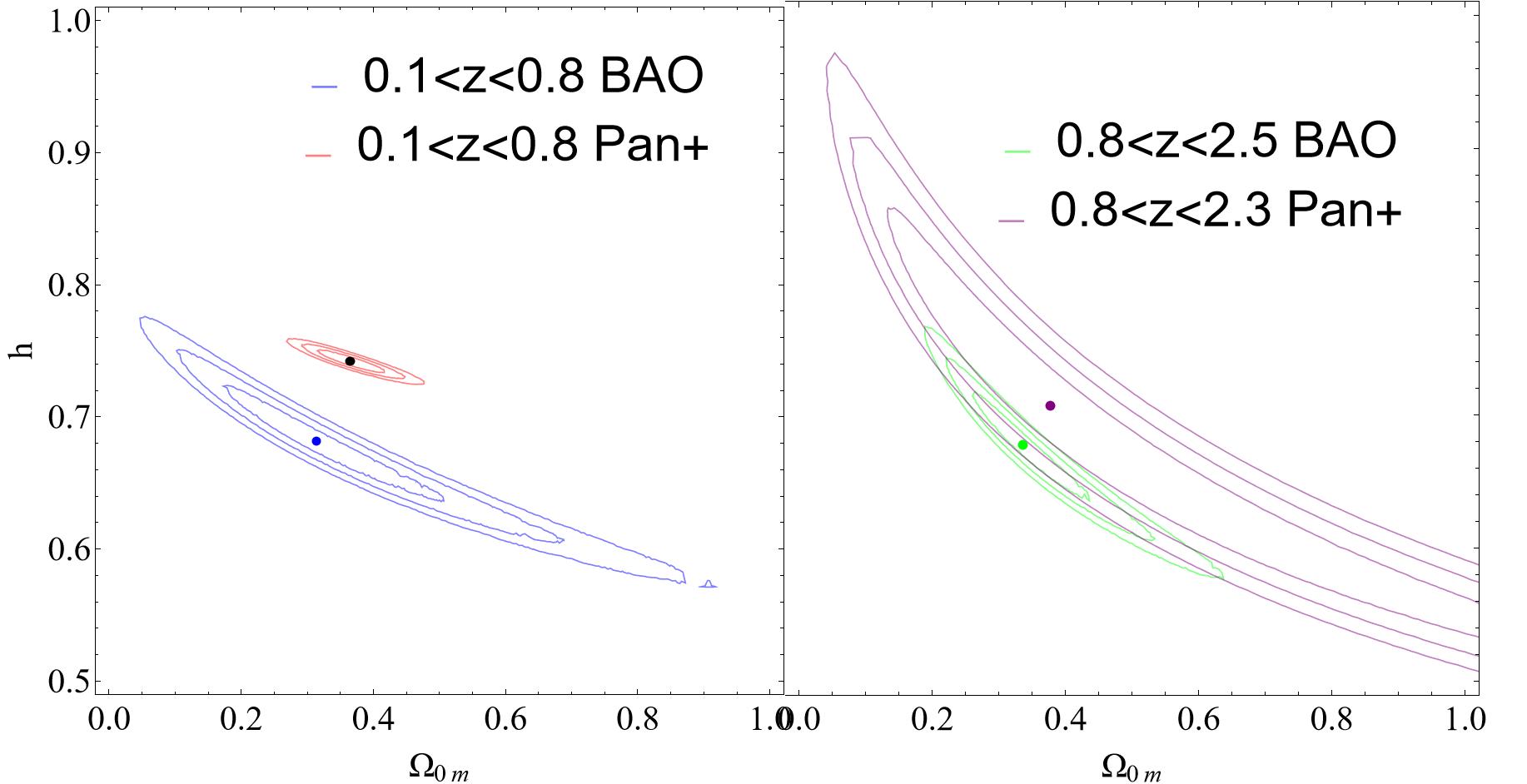}
\caption{The consistency between BAO and Pantheon+ data for the low (left) and high (right) redshift bins. Notice that the discrepancy is significantly larger for the low redshift bin.}
\label{fig2}
\end{figure*}

\begin{figure*}[ht!]
\centering
\includegraphics[width=\textwidth]{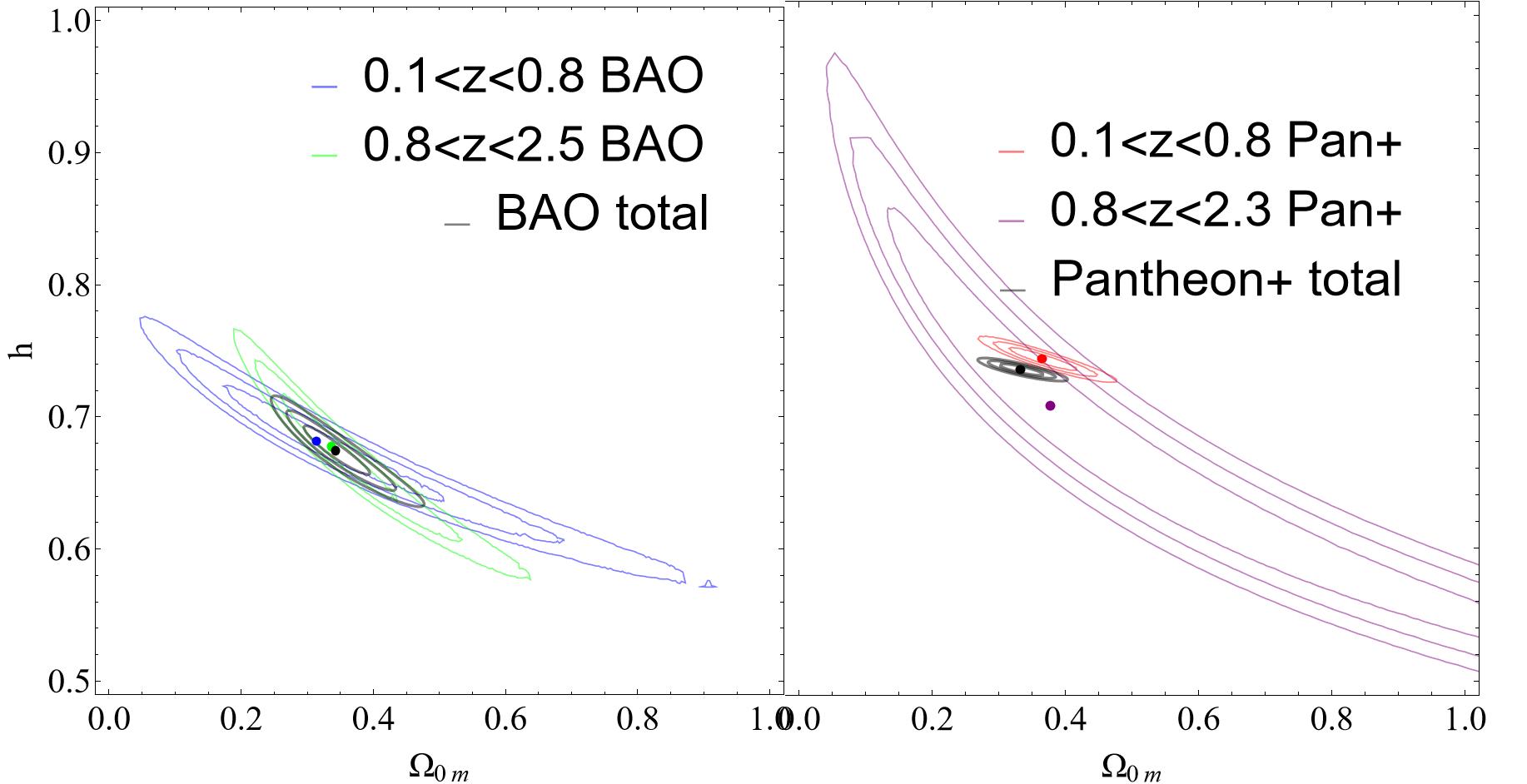}
\caption{The consistency between low and high redshift bins for BAO (left) and Pantheon+ (right) data. Notice that there is no evident inconsistency between the redshift bins in the context of each individual dataset. }
\label{fig3}
\end{figure*}

\begin{figure}[ht!]
\centering
\includegraphics[width=0.5\textwidth]{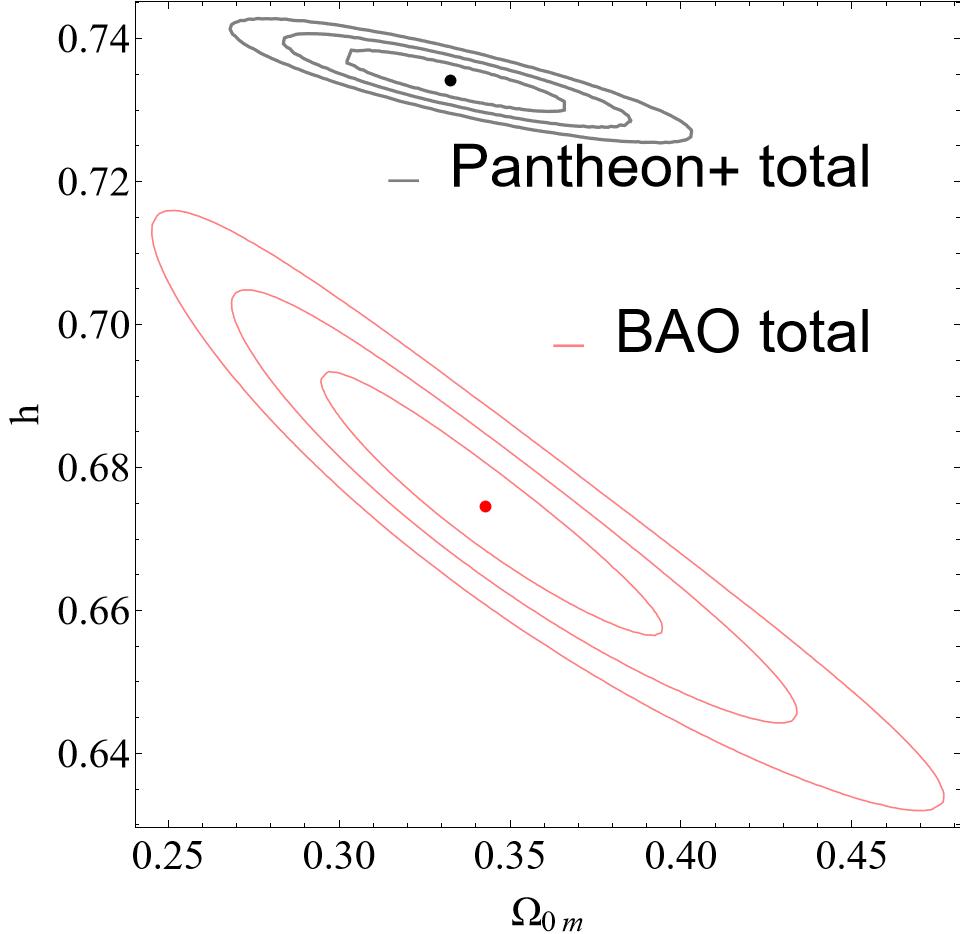}
\caption{The likelihood contours of \lcdm parameters obtained with the full BAO and Pantheon+ samples. The tension between the total BAO and Pantheon+ data sets in intermediate redshifts is apparent.}
\label{fig4}
\end{figure}

\begin{table*}
\centering
\caption{Best-fit cosmological parameters for two redshift bins within the \lcdm model.}
\label{tab:cosmo_params}
\begin{tabular}{cccc}
\hline
\textbf{Redshift Bin}             & $\boldsymbol{\Omega_{0m}}$ & $\boldsymbol{h}$ & $\boldsymbol{M_B}$    \\ \hline
First BAO Bin ($0.1 < z < 0.8$)   & $0.3 \pm 0.1$& $0.68 \pm 0.03$& -                   \\
Second BAO Bin ($0.8 < z < 2.3$)  & $0.34 \pm 0.06$& $0.68 \pm 0.03$& -                   \\
First Pan+ Bin ($0.1 < z < 0.8$)  & $0.37^{+0.04}_{-0.06}$& $0.742^{+0.008}_{-0.01}$& $-19.21^{+0.02}_{-0.03}$\\
Second Pan+ Bin ($0.8 < z < 2.3$) & $0.38^{+1}_{-0.2}$& $0.71^{+0.1}_{-0.2}$& $-19.20^{+0.7}_{-0.4}$\\
Total BAO                         & $0.34 \pm 0.03$& $0.67 \pm 0.01$& -                   \\
Total Pantheon+                   & $0.33 \pm 0.02$& $0.73 \pm 0.01$& $-19.25 \pm 0.03$\\ \hline
\end{tabular}
\end{table*}

The luminosity distance modulus for the model is as defined in equation (\ref{mu}) and the luminosity distance is as defined in equation (\ref{dl}), with the adjustment of using $H$ as defined in equation (\ref{wcdmH}). 

Similarly, we construct the distance modulus residual for the \lscdm model \cite{Akarsu:2021fol} with best fit parameter values as obtained in \cite{Akarsu:2023mfb}. The best fit parameter values we use for the \lscdm model are taken as $z_* = 1.72$, $\Omega_{m,\Lambda sCDM}=0.2646$ and $h_{0,\Lambda sCDM}=0.731$. These are taken from \cite{Akarsu:2023mfb}, specifically from Table 1 last column.

\section{Results: The Hubble Tension in Redshift Bins}
\label{sec:results}

The focus of our analysis is the level of mismatch of the distance moduli measured using BAO with the corresponding distance moduli measured with SnIa at various redshift bins. This is mainly demonstrated in Fig. \ref{fig1} which shows the distance moduli residuals with respect to \plcdm of the following:
\begin{itemize}
    \item The BAO data of the Table \ref{Dmrd} compilation (blue points). The latest DESI data are the green points shown separately for comparison with earlier points. The full BAO data are binned in wide redshift bins in the right panel.
    \item The binned Pantheon+ data (red points). These are binned in wide redshift bins in the right panel.
    The full Pantheon+ distance moduli are also shown in the grey background. The bin corresponding to $z<0.01$ has a much higher residual distance modulus compared to other bins due to the volumetric redshift scatter bias \cite{Perivolaropoulos:2023iqj}.
    \item The \plcdm model with uncertainties (upper band around zero).
    \item The Pantheon+ best fit \lcdm model (lower band around -0.2).
    \item The $wCDM$ model designed to interpolate between the SH0ES $H_0$ and the Planck18 angular diameter distance to recombination consistent with the CMB angular power spectrum data (green line)
    \item The best fit \lscdm model with respect to Pantheon+, BAO, SH0ES and CMB data with best fit discussed in the previous section.
\end{itemize}
The following comments can be made on the results shown in Fig. \ref{fig1}
\begin{itemize}
    \item There is a clear overall tension between the distance moduli measured by BAO and SnIa data. The distance moduli \plcdm residuals measured by BAO appear to be systematically higher than the distance moduli residuals measured by SnIa calibrated by Cepheids but lower compared to \plcdm expecially at $z>1$. Similarly, the model independent binned measured SnIa distance moduli appear to be slightly higher than the expected with respect to the best fit Pantheon+ \lcdm residual band (the lower stripe in the figure).
    \item This tension is less significant for redshift $z>1$ where the SnIa distance moduli residuals with respect to \plcdm increase while the BAO distance moduli residuals appear to decrease. This reduction of the tension at high $z$ may be viewed as a hint for an $H(z)$ deformation with respect to \plcdm which however by itself is not sufficient to fully resolve the Hubble tension. This deformation is better expressed by the best fit \lscdm model than by the best fit $wCDM$ model.
    \item Both $H(z)$ deformation models, $wCDM$ and \lscdm interpolate successfully between the \plcdm distance modulus to recombination and the Pantheon+ \lcdm distance moduli measured by the distance ladder calibrated SnIa at low $z$. However they can not fit simultaneously the BAO and SnIa distance moduli measured at intermediate redshifts since these are inconsistent with each other.
    \item The \lscdm model, due to its high $z$ $H(z)$ transition, appears to provide a better fit to the more constraining SnIa data than the smooth $wCDM$ best fit model. This seems to explain its overall improved quality of fit to the data compared to other $H(z)$ deformation models that attempt to resolve the Hubble tension.
\end{itemize}

The consistency level between BAO and Pantheon+ SnIa data in the context of \lcdm is demonstrated also in Fig. \ref{fig2}. The \lcdm $h$-\omom likelihood contours are shown for both BAO and SnIa for two large redshift bins $z\in [0.1,0.8]$ and $z\in [0.8,2.5]$. Clearly the best fit parameter values are in more than $4\sigma$ tension in the low redshift bin while the tension reduces to less than $2\sigma$ in the high redshift bin.

The internal consistency between the high and low redshift bins of the BAO and Pantheon+ SnIa samples is tested in Fig. \ref{fig3} where we show the parameter likelihood contours for the two bins of the BAO-Pantheon+ samples and for full samples. For the BAO sample the two redshift bins are clearly consistent with each other and with the full BAO dataset while for the Pantheon+ sample there is a mild $2\sigma$ tension between the low redshift bin and the full sample in the direction of $h$. This is also shown in the best fit parameter values with uncertainties shown in Table \ref{tab:cosmo_params}. Finally, the tension between the full BAO and full Pantheon+ SnIa data is more clearly demonstrated in Fig. \ref{fig4}.

\section{Conclusion and Discussion}
\label{sec:conclusion}

We have explored the level of consistency 
between the distance moduli of an extensive up to date compilation of BAO and the Pantheon+ SnIa data as a function of redshift. Using a direct comparison of the measured luminosity distance moduli in redshift bins as well as a comparison of the \lcdm best fit parameter values in low and high $z$ redshift bins we demonstrated the  level of tension between the BAO and SnIa at various redshift bins. This discrepancy is mainly due to the well known inconsistency of the calibrators used in each sample (recombination sound horizon vs distance ladder calibration of SnIa with Cepheids) known as the Hubble tension. Since we have used data at intermediate redshifts we were able to identify the redshift dependence of this tension as expressed through BAO and SnIa data. In this context our analysis may be viewed as a redshift tomography of the Hubble tension.

We found that the Hubble tension has more prominent presence in relatively lower redshifts $z\in [0.1,0.8]$. This effect is not only due to the higher precision of both the BAO and SnIa data in this redshift range. It is also due to a trend of the best fit values of both distance moduli and \lcdm best fit parameters to become more consistent between BAO and SnIa at higher redshift bins ($z\in [0.8,2.5]$. This improved consistency between BAO and SnIa data at higher redshift compared to lower redshifts in the context of distance moduli and \lcdm best fit parameters, may also be interpreted as a high $z$ deformation of $H(z)$ with respect to \lcdm . 

Even though, this deformation can be well expressed by properly designed models like \lscdm, it is clear from our results that it can not by itself provide good fit to all data. This is due to the calibrator induced inconsistency between the BAO and Pantheon+ SnIa data which is more prominent at the redshift range $z\in [0.1,0.8]$ \footnote{This inconsistency between BAO and SnIa distance moduli was first pointed out in Ref. \cite{Pogosian:2021mcs} using older BAO and SnIa data, even though no redshift tomography was implemented there.}.

While there may be correlation among some of the BAO points of our sample, this covariance is in most cases not available in the literature. Thus, by including most available BAO data, we chose to maximize information rather than independence.  This approximation may have introduced some bias in our results. Since our sample is not very extensive we estimate that this bias is subdominant and does not change the main features of our results. 

Interesting extensions of the present analysis include the consideration of additional data like Cosmic Chronometers or multimessenger standard siren data for the determination of the distance moduli in the redshift range considered, This direct comparison with the measured BAO and SnIa measured distance moduli could provide hints about the accuracy of the assumed calibrators (sound horizon vs Cepheid based distance ladder)

{\bf Numerical analysis files:} The Mathematica v13 notebooks and data that lead to the construction of the figures of the paper may be downloaded from \href{https://github.com/DimitriosBousis/Hubble-tension-tomography}{this url.}

\section*{Acknowledgements}
We thank Adrià Gómez-Valent, Sunny Vagnozzi, Santi Avila  and Eoin Colgain for useful comments.
This article is based upon work from COST Action CA21136 - Addressing observational tensions in cosmology with systematics and fundamental physics (CosmoVerse), supported by COST (European Cooperation in Science and Technology). This project was also supported by the Hellenic Foundation for Research and Innovation (H.F.R.I.), under the "First call for H.F.R.I. Research Projects to support Faculty members and Researchers and the procurement of high-cost research
equipment Grant" (Project Number: 789).

\section{Appendix}
\label{Appendix}

Here we explain the way in which we obtained the $D_M/r_d$ measurements of Table \ref{Dmrd} for the datapoints with *:

For the measurements 1 and 9, \cite{BOSS:2016goe} provides the following measurements of $D_A(z)*(r_{s,fid}/r_s)$: $D_A(0.32)*(r_{s,fid}/r_s)=956 \pm 28 Mpc$ and $D_A(0.59)*(r_{s,fid}/r_s)=1421 \pm 23
Mpc$. Since $D_M=D_A(z+1)$,to obtain the $D_M(z)/r_d$ values shown on the table we simply need to multiply by $(r_s/r_{s,fid})$ to obtain the measurement of $D_A$ and then to multiply by $((z+1)/r_d)$. The fiducial value $r_{s,fid}$ in \cite{BOSS:2016goe} is $r_{s,fid}=147.66$ and we use $r_d=147.18$ and $r_s=147.09$. We do the same to obtain the uncertainty of the measurement.

For the measurements 4, 10 and 15, \cite{Blake:2012pj} provides the following measurements of $D_A(z)$: $D_A(0.44)=1205 \pm 114 Mpc$, $D_A(0.60)=1380 \pm 95 Mpc$ and $D_A(0.73)=1534 \pm 107 Mpc$. To obtain $D_M/r_d$ we simply need to multiply by $((z+1)/r_d)$. We do the same to obtain the uncertainty of the measurements. \cite{Seo:2012xy} (measurement 7) and \cite{BOSS:2014hwf} (measurement 30) also provide measurements of $D_A$ and thus we follow the same steps. \cite{Seo:2012xy} provides the following: $D_A(0.54)=1411 \pm 65 Mpc$. \cite{BOSS:2014hwf} provides the following: $D_A(2.34)=1662 \pm 96 Mpc$.

For measurements 12 and 20, \cite{Sridhar:2020czy} provides the following: $D_A(0.697)=(1499 \pm 77 Mpc) (r_d/r_{d,fid})$ and $D_A(0.874)=(1680 \pm 109 Mpc) (r_d/r_{d,fid})$. Therefore, to obtain measurements of $D_M/r_d$ we need to multiply the results of $D_A$ by $((z+1)/rd$. We do the same to obtain the uncertainty of the measurement. In \cite{Sridhar:2020czy} the fiducial value of $r_d$ is $r_{d,fid}=147.21$.

%\subsection{Proof for the uncertainty equation \ref{UncDm} for the BAO data}
%We use error propagation to calculate the resulting total uncertainty.

%\begin{multline}
%\sigma_i=\sqrt{\left( \frac{\partial}{\partial D_M^{obs}(z_i)}\Delta_{\mu}(z_i) \right)^2 \cdot \sigma_{ D_M^{obs}}^2(z_i)}= \\
%\\
%\left( \frac{\partial}{\partial D_M^{obs}(z_i)}\Delta_{\mu}(z_i) \right) \cdot \sigma_{ D_M^{obs}}(z_i)= \\
%\\
%\frac{\partial}{\partial D_M^{obs}(z_i)}\left( 5log(\frac{D_M^{obs}(z_i)}{D_M^{th}(z_i)}) \right) \cdot \sigma_{ D_M^{obs}}(z_i)= \\
%\\
%\frac{5}{ln10}\frac{\frac{\partial}{\partial D_M^{obs}(z_i)}(D_M^{obs}(z_i)/D_M^{th}(z_i))}{D_M^{obs}(z_i)/D_M^{th}(z_i)} \cdot \sigma_{ D_M^{obs}}(z_i) = \\
%\\
%\frac{5}{ln10}\frac{1/D_M^{th}(z_i))}{D_M^{obs}(z_i)/D_M^{th}(z_i)} \cdot \sigma_{ D_M^{obs}}(z_i)
%\end{multline}
%\section{Data Used in the Analysis}
%\label{sec:Appendix_A}

\raggedleft
\bibliography{References}

\end{document}